\documentclass[12pt]{article}
\makeatletter\@addtoreset{equation}{section}\makeatother

\newtheorem{Tm}{Theorem}[section]
\newtheorem{Rk}{Remark}[section]
\newtheorem{Lm}{Lemma}[section]
\newtheorem{Co}{Corollary}[section]
\newtheorem{Df}{Definition}[section]
\newtheorem{Pn}{Proposition}[section]

\begin{document}
\begin{center}
\begin{Large}
{\bf  Classical Limits of Euclidean Gibbs States for Quantum
  Lattice Models}
\end{Large}
\end{center}

\vskip.8cm
\begin{center}
\begin{Large}
 Sergio Albeverio\\
\end{Large}
\begin{large}
 Abteilung f\"ur Stochastik, \\
  Institut f\"ur Angewandte Mathematik, Universit\"at Bonn,\\
        D 53115 Bonn
        (Germany); \\
        Forschungszentrum BiBoS, Bielefeld
(Germany);\\
        SFB 237 (Essen--Bochum--D\"usseldorf) (Germany); \\
        CERFIM and USI, Locarno (Switzerland) \\
        e-mail albeverio@uni--bonn.de
\end{large}
\vskip0.4cm
\begin{Large}
      Yuri Kondratiev \\
 \end{Large}
 \begin{large}
 Abteilung f\"ur Stochastik,\\
   Institut f\"ur Angewandte Mathematik, Universit\"at Bonn,\\
     D 53115 Bonn
        (Germany); \\
     Forschungszentrum BiBoS,    D 33615 Bielefeld (Germany); \\
        Institute of Mathematics, Kiev (Ukraine)\\
        e-mail kondratiev@uni-bonn.de
\end{large}
\vskip0.4cm
\begin{Large}
      Yuri Kozitsky \\
\end{Large}
\begin{large}
        Institute of Mathematics, Maria Curie-Sklodowska University\\
        PL 20-031 Lublin (Poland);\\
     Institute for Condensed Matter Physics, Lviv (Ukraine)\\
        e-mail jkozi@golem.umcs.lublin.pl
\end{large}
\end{center}

\def\cit#1{\cite{[#1]}}
\def\kasten{\hfil\vrule height6pt width5pt depth-1pt\par }
\date{}
\def\C{{C\!\!\! C}}
\def\R{{I\!\! R}}
\def\Rr{{I\!\! R}^{\delta}}
\def\N{{I\!\! N}}
\def\Z{{Z\!\!\! Z}}
\def\Zd{{Z\!\! Z}^d }
\def\pp{\partial}
\def\ds{\displaystyle}
\def\half{{1 \over 2}}
\def\lra{\longrightarrow}
\def\Ra{\Rightarrow}
\def\1{{1\!\!1}}
\def\Tri{\triangle}
\def\a{\alpha}
\def\b{\beta}
\def\g{\gamma}
\def\d{\delta}
\def\e{\epsilon}
\def\i{\iota}
\def\k{\kappa}
\def\t{\tau}
\def\l{\lambda}
\def\om{\omega}
\def\s{\sigma}
\def\vp{\varphi}
\def\ve{\varepsilon}
\def\vt{\vartheta}
\def\D{\Delta}
\def\G{\Gamma}
\def\L{\Lambda}
\def\Om{\Omega}
\def\AC{{\cal A}}
\def\BC{{\cal B}}
\def\CC{{\cal C}}
\def\BR{{\rm B}}
\def\Bq{{\rm B}^{{\rm qc}}}
\def\DC{{\cal D}}
\def\EC{{\cal E}}
\def\FC{{\cal F}}
\def\GC{{\cal G}}
\def\GCq{{\cal G}^{\rm qc}}
\def\GCc{{\cal G}^{\rm cl}}
\def\HC{{\cal H}}
\def\Hb{{\cal H}_{\beta}}
\def\IC{{\cal I}}
\def\KC{{\cal K}}
\def\LC{{\cal L}}
\def\MC{{\cal M}}
\def\MC1{{\cal M}_1}
\def\NC{{\cal N}}
\def\PC{{\cal P}}
\def\XC{{\cal X}}
\def\XoB{{\cal X}({\cal B})}
\def\lv{\left\vert}
\def\rv{\right\vert}
\def\bb#1{ {\lv #1 \rv} }
\def\bq#1{ {\lv #1 \rv}^2 }
\def\BE#1{ {\left\Vert #1 \right\Vert} }
\def\BQ#1{ {\left\Vert #1 \right \vert}^2 }
\def\Ti#1{ {\tilde{#1}}}
\def\TR{{\rm trace}}
\def\ti{\; \times \!\!\!\!\!\!\!\! \mathop{\phantom\sum}}
\def\dop{\dot +}
\def\C{\hbox{\vrule width 0.6pt height 6pt depth 0pt \hskip -3.5pt}C}
\def\ope{\;
\lra_{\!\!\!\!\!\!\!\!\!\!\!\!\!_{\hbox{$_{n\to\infty}$}}} \;}
\def\SB{\bf S}
\def\be{\begin{equation}}
\def\th{\theta}
\def\ee{\end{equation}}
\def\SC{\cal S}
\def\beq{\begin{eqnarray}}
\def\eeq{\end{eqnarray}}
\def\bw{\bar{w}}
\def\p{2^{\delta}}
\def\2m{2^{-\delta}}
\def\k{\kappa}
\def\FB{{\cal F}^{(e)}_{\beta,\delta}}
\def\ra{\rightarrow}
\def\oo{[\beta_{0}^{-}, \beta_{0}^{+}]}
\def\bk{\bar{\kappa}}
\def\Hos{{\cal H}^{\rm osc}}
\def\Ho{{\cal H}^{\rm osc}}
\def\Hoj{{\cal H}^{\rm osc}_j }
\def\HoL{{\cal H}^{\rm osc}_{\Lambda_{n,j}} }
\def\Lf{{\cal L}_{{\bf fin}}}
\def\Lan{{\Lambda}_{n,j}}
\def\Ln{{\cal L}_n}
\def\Lal{{\Lambda}_{n-l,s}}
\def\La{{\Lambda}_{n-1,s}}
\def\Th{{\cal T}_{{\bf hier}}}
\def\z{\zeta}
\def\CN{{\cal N}}
\def\sta{{\stackrel{\rm def}{=}}}
\def\Lf{{\cal L}_{\rm fin}}
\pagenumbering{arabic}

\begin{abstract}
Models of quantum and classical particles on the $d$--dimensional
lattice $\Zd$
with pair interparticle interactions are
considered. The
classical model is obtained from the corresponding quantum one when
the
reduced physical mass of the
particle $m = \mu /\hbar^2  $ tends
to infinity. For these models,
it is proposed to define the convergence of the Euclidean Gibbs
states, when $ m\ra +\infty$, by the weak convergence of
the corresponding local
Gibbs specifications, determined by conditional Gibbs measures.
In fact it is proved that all conditional
Gibbs measures of the quantum model weakly converge to the conditional
Gibbs measures of the classical model.
A similar convergence of the periodic Gibbs
measures and, as a result, of the order parameters, for
such models with pair interactions possessing the
translation invariance,
has also been proven.

\end{abstract}

\section{Introduction}

We consider a system of interacting particles performing one--
dimensional
oscillations around their equilibrium positions which form a $d$--
dimensional
lattice $\Z^d \subset \R^d$. This system serves as a base for two models. The
first one
is a quantum mechanical
model described by the following formal Hamiltonian
$$
H= -\frac{1}{2m}\sum_{k\in \Z^d}\Delta_k + W ,
$$
where $\Delta_k $ stands for the Laplacian (in our case it is simply
$d^2 /dx_k^2$, $x_k \in \R$) and
$m>0$ is the reduced physical mass of the particle, which
one obtains dividing the physical mass by $\hbar^2 $.
The first term
corresponds to the kinetic energy of the particles,
the second one describes their potential energy including the
crystalline field as well as the energy of the interparticle
interaction. We consider the model where this interaction is
pairwise. As a reference model, we choose the model of noninteracting
harmonic oscillators. Hence the formal Hamiltonian now is
\be
\label{N1}
H= \sum_{k} H^{(0)}_k + \sum_{k }U_k (x_k ) + \half
\sum_{j,k }J_{jk}x_j x_k ,
\ee
where all sums are taken over the lattice $\Z^d $.
The first term is the Hamiltonian of the reference model, i.e.
the sum of the Hamiltonians of identical
harmonic oscillators. Each $H_k^{(0)}$ is defined in the complex Hilbert space
$L^2 (\R , dx_k )$ (see e.g. \cit{BeS}) and reads
\be
\label{2}
H_k^{(0)} = -\frac{1}{2m} \Delta_k + \frac{1}{2}x_k^2 .
\ee
The second term in (\ref{N1})
contributes to the crystalline field
making it to be anharmonic. The third term describes
the interaction between the particles. For the matrix
$J_{jk}$, we assume that there exists $r>0$ such that
$J_{jk} =0$ whenever the Euclidean distance $\vert j-k \vert $
exceeds this $r$.
We also suppose that
\be
\label{c}
c \ \sta \ {\rm sup}_j \sum_{k} \vert J_{jk} \vert < \infty.
\ee
For the functions $U_k : \R \ra
\R$, we assume that all of them are continuous and
that the following estimate holds,
for all $k\in \Z^d$ and all $x\in \R$,
\be
\label{Qu2}
U_k (x) \geq \half \tilde{c} x^2 + b, \ \ \ \tilde{c}>\max\{c-1 , 0\}.
\ee
Here the parameter $c$ is defined by (\ref{c}) and
 $b\in \R$.
The second model is a classical version of the model described above.
Its formal Hamiltonian is
\be
\label{Cl1}
H^{{\rm cl}} = \sum_{k }\left(\frac{1}{2}x_k^2 + U_k (x_k )\right)  +
\half
\sum_{j,k }J_{jk}x_j x_k ,
\ee
which means that in this case
only
the potential energy $W$ is
taken into account (see \cit{AH--K}). Heuristically
the latter Hamiltonian
may be obtained from the quantum one (\ref{N1}) by passing to the
limit
$m\ra +\infty$.

The aim of this work is to study the possible convergence of
the Gibbs states of the quantum model
to the Gibbs states of the classical model.
It should be remarked that the convergence of certain Green functions,
describing the Gibbs states, in similar models with a special choice
of the interaction potentials has already been proved  in
\cit{AH--K}.

For the quantum lattice models, the Gibbs states
are constructed as positive functionals on the algebras of
observables (see e.g. \cit{BrR}, \cit{KL}), in  contrast to the case
of classical models where they are built by means of conditional
probability
distributions (see \cit{Do}), which form the so called
Gibbs specifications
(see  \cit{Ge}), as measures which solve
the equilibrium (Dobrushin--Lanford--Ruelle)
equations.
But for the quantum models with
unbounded operators, which we consider here,
the algebraic approach does not allow to construct such states
of infinite systems.
In 1975 \cit{AH--K}
an approach to the construction
of Gibbs states of quantum lattice models
has been initiated. This approach uses
the integration theory in path spaces
(see also \cit{AH--K1},
\cit{BaK1}--\cit{BaK3}, \cit{GK}, \cit{KL}, and \cit{Si2}).
 Here the state of an
infinite system at temperature $\b^{-1}$
is defined by a
probability measure $\nu_\b $ on a certain space $\Om_\b $. As in
the case
of classical systems, this measure solves the equilibrium equation and
the Gibbs state as a positive functional may be
reconstructed, analogously to the Euclidean quantum field
theory,
  by means of the moments of the measure $\nu_\b $, which here are
the temperature Green functions. That is why the measure $\nu_\b$ is known as
{\it the Euclidean Gibbs state} of the quantum model. In the frames of such {\it
an Euclidean} approach, it has become possible to develop substantially the
theory of Gibbs states in quantum models with unbounded operators. The
additional advantage of this Euclidean approach is that the Gibbs states in
quantum and in classical models may be considered in one and the same setting.
This setting allows to define more precisely what does it mean that quantum
Euclidean Gibbs states converge to corresponding classical Gibbs states. For a
given model at given temperature, a family of Euclidean Gibbs states is the
family of solutions of the equilibrium equation defined by the local Gibbs
specification. This family may consist of several elements. The same holds also
for the corresponding classical model. The mentioned advantage lies in the fact
that both quantum and classical Gibbs states may be defined as measures on one
and the same space. The best possible way to study the convergence being
discussed here is to show how each element of the family of Euclidean Gibbs
states converges to the corresponding classical Gibbs state. But, even in the
case where the cardinalities of the families of quantum and classical Gibbs
states are equal, it would be very difficult to show the convergence of
Euclidean Gibbs states except perhaps for very special cases.
 This would be all the more so in the case where
these cardinalities are different (which implies that some bifurcations
of the quantum states at certain values of the reduced mass $m$ take
place). In this paper, we propose to define the
convergence of Euclidean Gibbs states of a quantum model to
corresponding Gibbs states of a classical model in terms of the
convergence of their local Gibbs specifications. Such a convergence,
as it is shown below, holds in some sence
at all values of the inverse temperature, which
means in particular
it holds even when the mentioned cardinalities are different. The
possibility of families having different cardinalities, i.e. of
phase transitions to take place, for sufficiently large values of
the inverse temperature $\b$
follows from the results of our recent
works (see \cit{AKK1}, \cit{AKKR}, \cit{BaK2}
and references therein). For background concerning the main features
of our technique we refer to
\cit{AKRT} -- \cit{AKKR}, \cit{BaK1}, \cit{BaK2}, \cit{BaK3},
\cit{GK}.

\section{Euclidean Gibbs States}
Let $\LC$, $\Lf$ denote the set of all, respectively of all finite,
subsets of $\Zd$.
For certain value of the inverse temperature $\b >0$,
 we consider the space of
continuous functions ({\it temperature loops})
taking equal values at the endpoints of the interval $[0,\b]$
$$
C_\b  \ \sta \ \{ \om \in C([0,\b] \ra \R ) \ \vert \
 \om (0) = \om (\b) \},
$$
equipped with the norm
\be
\label{norm}
\vert \om \vert_\b \ \sta \ {\rm sup }\{\vert \om (\t ) \vert :\
 \t \in [0, \b] \},
\ee
and with the usual Banach space structure. For $\L \in \LC$, we put
\be
\label{3aa}
\Om_{\b , \L } \  \sta \ \{ \om_\L = (\om_k )_{k\in \L} \ \vert \
\om_k \in C_\b  \}, \ \ \  \ \Om_\b \ \sta \ \Om_{\b ,
\Z^d }.
\ee
$\Om_{\b , \L}$
will be called {\it the temperature loop spaces} (TLS), their elements
are {\it the configurations of
the temperature loops} (at the inverse
temperature $\b$ and for the domain $\L$).
The TLS  $\Om_{\b , \L}$  may be
equipped with the product topology and with the
$\s$--algebra ${\cal B}_{\b , \L}$
generated by the cylinder
subsets of $\Om_{\b , \L} $,
 see e.g. \cit{AKRT}, \cit{Ge}, \cit{PaY1}, \cit{PaY2}.

In order to have the collection of all TLS $\{\Om_{\b , \L} , \L \in
\LC \}$
ordered by inclusion, one may introduce the following embedding
mappings. For $\L \subset \L'$, we put $\om_\L  \mapsto \om_\L \times
0_{\L'\setminus \L} \in \Om_{\b , \L'}$. Here $0_\L $ is the zero
configuration in $ \Om_{\b , \L}$, and
$$
\om_\L \times \xi_{\L' \setminus \L} \ \sta \ \z_{\L'}  , \ \L
\subset \L',
$$
means the configuration in $ \Om_{\b , \L'}$ such that $\z_k = \om_k$
for $k\in \L$, and $\z_k = \xi_k $ for $k\in \L' \setminus \L$.
Having in mind such embeddings, we shall consider every
configuration $\om_\L $ as an element of all TLS $\Om_{\b , \L'}$,
with $\L\subset \L'$.
Along with these embeddings, we define the projections
$$
\om_\L \mapsto (\om_\L )_{\L'} \ \sta \ (\om_k )_{k\in \L \cap \L' },
$$
as a configuration in $ \Om_{\b , \L'}$ such that $\om_k =0$,
for $k\in\L' \setminus \L$. Obviously, then
$(\om_\L )_{\L'}$ is the zero configuration if $ \L \cap \L' =
\emptyset$.

It is easily seen that under the assumptions made regarding
the potentials $U_j $ and $J_{jk}$ the following expression
\beq
\label{A1}
E_{\b , \L}(\om_\L \vert \z ) \ & \sta & \ \sum_{j  \in \L }
\int_{0}^{ \b}U_j (\om_j (\tau ) ) d\tau
 +  \half \sum_{j,k \in \L}
\int_{0}^{\b}J_{jk} \om_j (\tau ) \om_k (\tau ) d\tau \nonumber \\
& + & \sum_{j\in \L ,
k \in \L^c } \int_{0}^{\b }J_{jk} \om_j (\tau )\z_k (\tau)) d\tau ,
\ \ \ \L \in \Lf ,
\eeq
defines a continuous function
$
 E_{\b , \L}(. \ \vert \z ): \ \Om_{\b , \L} \longrightarrow \R .$

The space of temperature loops $C_\b $
 may
naturally be embedded into the real Hilbert
space $\Hb \sta L^2 ([0, \b])$ with
the scalar product
\be
\label{s2}
(\om , \xi )_\b
=  \int_{0}^{\b }\om (\t) \xi (\t) d\t .
\ee
Let ${\cal H}_{\b , j }$, $j\in \Zd$ be the $j$--th
copy of $\Hb$. For $\L \in \Lf$, we put
\beq
\label{HH4}
{\cal H}_{\b , \L} & \sta  & \bigoplus_{j \in \L} {\cal H}_{\b , j}
 =  \{ \om_\L = (\om_j )_{j\in \L} \
\vert\ \om_j \in {\cal H}_{\b , j } \}.
\eeq
For short, we omit in the sequel a subscript like $j$
when this does not cause any ambiguities.

The scalar product in the Hilbert space ${\cal H}_{\b , \L}$ is
\be
\label{3ab}
(\om_\L , \xi_\L )_{\b , \L } = \sum_{j \in \L} (\om_j , \xi_j )_\b ,
\ee
and for all $\L \in \Lf$, one has
$
\Om_{\b , \L } \subset {\cal H}_{\b , \L}.$

Let us consider the following strictly positive trace
class operator on $\Hb$
\be
\label{Sm}
S_{\b}(m) = (-m\D_\b + 1)^{- 1},
\ee
where $\D_\b $ stands for the Laplace operator in $L^2 ([0, \b])$
 and $m$
is, as above, the reduced physical mass of the particle.
Then one can define the Gaussian
measure $\g_\b^{(m)} $ on
$\Hb $ which has zero mean and $S_\b (m) $
as a covariance operator. The measure $\g_\b^{(m)} $ is uniquely determined
by its Fourier transform
\be
\label{5}
\int_{\Hb}\exp(i(\vp,\om)_{\b})\g_\b^{(m)} = \exp (-
\frac{1}{2}(\vp, S_{\b} (m)\vp )_{\b}) , \ \ \vp\in\Hb .
\ee
Actually, the set of continuous loops is a set of full measure
(i.e. $\g_\b^{(m)} (C_\b )=1 $), and the measure $\g_\b^{(m)}$
corresponds to the oscillator bridge process of length $\b$
\cit{Si2}. Then the product
 measure
\be
\label{0c}
\g_{\b,\L}^{(m)}(d\om_\L ) \ \sta \ \prod_{j\in \L} \g_\b^{(m)}(d\om_j ),
\ \ \L \in \Lf,
\ee
is a measure defined on ${\cal H}_{\b , \L}$
and supported on $\Om_{\b , \L}$.

Now let the subset $\L \in \Lf$ be fixed.
For the model considered, the Gibbs measure in $\L$, subject to
a configuration $\z\in \Om_{\b }$, is
\be
\label{9}
\nu_{\b , \L}^{(m)}(d\om_{\L}\vert\zeta)
 \ \sta \  \frac{1}{Z_{\b,\L}(\zeta)}\exp\{-
E_{\b,\L}(\om_{\L}\vert \zeta) \} \g_{\b,\L}^{(m)}(d\om_\L ),
\ee
defined, as $\g_{\b , \L}^{(m)}$, on ${\cal H}_{\b , \L}$ and supported
on the space $\Om_{\b , \L}$. Here
\be
\label{11}
Z_{\b , \L}(\zeta) \ \sta \ \int_{\Om_{\b , \L}}\exp\{ - E_{\b ,
\L} (\om_{\L}\vert \zeta)\}\g_{\b,\L}^{(m)}(d\om_\L ),
\ee
is the finite volume partition function subject to the
external boundary condition $\zeta_{\L^{c}}$.
The conditions  (\ref{c}) and (\ref{Qu2}) imposed on the potentials
$J_{jk}$ and $U_k$
provide that the function $\exp\{ - E_{\b , \L} (\om_\L \vert \z )\}$
is $\g_{\b , \L}^{(m)}$--integrable, thus the objects introduced in
(\ref{9}),
(\ref{11}) are well--defined.

For $B\in {\cal B}_\b \sta {\cal B}_{\b , \Z^d}$, let
$\1_B $ be the indicator function of
$B$. Introduce the
family of probability kernels $\{ \pi_{\b , \L }^{(m)} \ \vert \ \L \in \LC
\}$
\be
\label{12}
\pi_{\b , \L}^{(m)} (B\vert \zeta ) \ \sta \ \int_{\Om_{\b , \L}} \1_{B}
(\om_\L \times \zeta_{\L^c })\nu_{\b , \L }^{(m)} (d\om_\L \vert \zeta ),
\ee
which satisfy the following consistency conditions
 (for more details see \cit{Ge}). For every $\L' \in \L$,
\beq
\label{13}
\pi_{\b , \L }^{(m)}\pi_{\b , \L' }^{(m)} (B\vert \zeta ) \ & \sta & \
\int_{\Om_\b }\pi_{\b , \L}^{(m)}(d\om \vert \zeta )\pi_{\b ,
\L' }^{(m)}(B\vert \om ) \nonumber \\
& = & \pi_{\b , \L}^{(m)}(B\vert \zeta).
\eeq
\begin{Df}
\label{1df}
A probability measure $\nu_\b $ on $(\Om_\b , {\cal B}_\b  )$ is
said to be an Euclidean Gibbs state of
the lattice model (\ref{N1}), (\ref{2}) at the inverse temperature
$\b$
if it satisfies the Dobrushin--Lanford--Ruelle (DLR)
equilibrium equation:
$$
\nu_\b \pi_{\b , \L}^{(m)} = \nu_\b ,
$$
 that is
\be
\label{14}
\int_{\Om_\b }\nu_\b (d\om)\pi_{\b , \L}^{(m)} (B\vert \om )  =
\nu_\b (B),
\ee
for all $\L \in \LC $ and $B \in {\cal B}_\b $.
\end{Df}
The class of all Euclidean Gibbs measures, i.e., the set of
solutions of (\ref{14}) is denoted ${\cal G}(\b)$.

\section{Quasiclassical States and Classical Limits}

In this section and in the subsequent one we present the
formulation of our results, referring to Section 5 for all proofs.
Given $\L \in \LC$, let us consider the subset of
$\Om_{\b , \L}$ consisting of constant trajectories,
that is
\be
\label{A2}
\Om^{{\rm qc }}_{\b , \L } \ \sta \ \{\om_\L \in \Om_{\b , \L } \
\vert  \ (\forall k\in \L ) \
(\forall \tau \in [0, \b ])  \ \ \om_k(\tau) = x_k \in \R  \},
\ee
which is isomorphic to $\R^\L $.
We also set
$$
\Om_\b \supset \Om^{{\rm qc }}_{\b , \Z^d}\ \sta \ \Om^{{\rm qc%
}}_{\b }
\cong \R^{\Z^d}.
$$

Further, for $\L \in \LC $, let ${\cal B}^{{\rm qc }}_{\b , \L}$
be the $\sigma$--algebra generated by the cylinder subsets
of $\Om^{{\rm qc }}_{\b , \L } $, which is isomorphic to the
corresponding $\s$-algebra ${\cal B} (\R^\L )$
generated by the cylinder subsets of $\R^\L $ but,
on the other hand, is a subalgebra of ${\cal B}_{\b , \L}$.
For every $B\in {\cal B}_{\b , \L}$, let
\be
\label{77}
{\rm C}(B) \ \sta \ B \cap \Om^{{\rm qc }}_{\b , \L } .
\ee
We will also write
\be
\label{A}
{\cal B}^{{\rm qc }}_{\b , \L} \ni
C\cong A \in {\cal B} (\R^\L ),
\ee
for the pair of subsets $C\in {\cal B}^{{\rm qc }}_{\b , \L}$,
 $A\in{\cal B} (\R^\L ) $ which are connected
by the isomorphism mentioned above. This means that they consist
of exactly those $\om_\L $ and $x_\L $, for which
$\om_j (\t ) = x_j $, for all $\t \in [0, \b]$ and $j\in \L$.

Consider the following Gaussian measures
\beq
\label{A6}
 \chi_{\b,  \L} (dx_\L ) & \ \sta \ & \prod_{j\in \L}
\chi_{\b} (dx_j ) ,  \ \ x_\L \in \R^\L ,\ \ \L \in \Lf, \\
\label{A7}
\chi_{\b} (dx_j ) & \ \sta \ & \sqrt{ \frac{\b}{2\pi }}
\exp\{ - \frac{\b}{2} x_j^2 \}dx_j ,\ \ x_j \in \R .
\eeq
For $\L \in \Lf$, let $\g_{\b , \L}^{{\rm qc }}$ be the
 Gaussian measure on $\Om_{\b , \L}$ such that for every
$ B\in {\cal B}_{\b , \L}$
\be
\label{Gcl}
\g_{\b , \L}^{{\rm qc }} (B) = \chi_{\b , \L} (A ) ,
\ee
where $A\cong {\rm C}(B)$, which is defined by (\ref{77}),
(\ref{A}). This means that
\be
\label{Gg}
\g_{\b , \L}^{{\rm qc }} (B) = \g_{\b , \L}^{{\rm qc }}
({\rm C} (B)) ,
\ee
i.e., it is supported on ${\cal B}^{{\rm qc }}_{\b , \L}$.

Making use of these measures we construct the conditional
Gibbs measures following the scheme (\ref{0c}) -- (\ref{14}).
As in (\ref{9}) we set
\be
\label{D}
\nu_{\b , \L}^{{\rm qc}}(d\om_{\L}\vert\zeta)
 \ \sta \  \frac{1}{Z_{\b,\L}^{{\rm qc }}(\zeta)}\exp\{-
E_{\b,\L}(\om_{\L}\vert \zeta) \} \g_{\b , \L}^{{\rm qc}}(d\om_\L ),
\ee
\be
\label{DZ}
Z_{\b,\L}^{{\rm qc }}(\zeta) \ \sta \ \int_{\Om_{\b,\L}}
\exp\{-
E_{\b,\L}(\om_{\L}\vert \zeta) \} \g_{\b , \L}^{{\rm qc}}(d\om_\L ),
\ee
which is defined on the same space as $\nu_{\b , \L}^{(m)}(. \vert\zeta)$
and with $\z \in \Om_\b $.
Further, (\ref{Gg}) implies
\be
\label{Ng}
 \nu_{\b , \L}^{{\rm qc }} (B) = \nu_{\b , \L}^{{\rm qc }}
({\rm C} (B)).
\ee
By means of the conditional Gibbs measures (\ref{D}), (\ref{DZ})
one can define
the family of probability kernels
$\{\pi^{{\rm qc}}_{\b , \L}(. \vert \z)\}$
 as well as the corresponding
Euclidean Gibbs states.
The family of such Euclidean Gibbs states will be denoted
${\cal G}^{{\rm qc}} (\b )$. The members of this family
will be called {\it quasiclassical} Gibbs states.

Now let us construct the Gibbs measures for the classical
model described by the Hamiltonian (\ref{Cl1}). To this end
we introduce a function
analogous to (\ref{A1}) which defines the interparticle interaction
in the classical model
\beq
\label{E}
I_{  \L}(x_\L \vert y ) \  \sta  \ \sum_{j  \in \L }
U_j (x_j )
 +  \half \sum_{j,k \in \L}
J_{jk} x_j x_k
 +  \sum_{j\in \L ,
k \in \L^c } J_{jk} x_j y_k  ,
\ \ \ \L \in \Lf ,
\eeq
where $y= (y_j )_{j\in\Z^{d}}\in \R^{\Z^d} $ determines the boundary
conditions outside $\L$ and plays here the same role as $\z$
in the case of Euclidean Gibbs measures. It is not difficult
to prove that
$I_\L (. \vert y)$ is a continuous function on $\R^\L $, $\L \in \Lf$.
A conditional Gibbs measure for the classical model is introduced
as follows
 \be
\label{A5}
\mu_{\b , \L} (dx_\L \vert \xi) = \frac{1}{Y_{\b , \L}(y)}
\exp \{ - \b I_{ \L} (x_\L \vert y )\}
\chi_{\b ,\L} (dx_\L ),
\ee
\be
\label{Y}
Y_{\b , \L}(y) = \int_{\R^\L}\exp \{ - \b I_{ \L} (x_\L \vert y )\}
\chi_{\b ,\L} (dx_\L ),
\ee
Like in the quantum case, the family
of conditional Gibbs measures
\newline $\{\mu_{\b , \L}(. \vert y) \vert \L \in
\Lf \}$ may be used to define the following probability kernels
\be
\label{clk1}
\rho_{\b , \L}(B\vert y ) \ \sta \ \int_{\R^{ \L}} \1_B
(x_\L \times y_{\L^c })\mu_{\b , \L } (dx_\L \vert y ), \ \ \
B\in {\cal B} (\R^{\Z^d }) ,
\ee
satisfying the consistency
condition analogous to (\ref{13}). The Gibbs states of the
classical model at given inverse temperature $\b$ are
understood in the sense of
Definition \ref{1df}.
They are the measures $\mu_\b $ on the space $\R^{\Z^d} $ which
satisfy the equilibrium equation
\be
\label{cl14}
\int_{\R^{\Z^d} }\mu_\b (dx)\rho_{\b , \L} (B\vert x )  = \mu_\b (B) ,
\ee
for all $\L \in \LC $ and $B \in {\cal B} (\R^{\Z^d })  $.
The family of Gibbs states for the classical model is denoted
${\cal G}^{{\rm cl}}(\b )$.

In the sequel we shall use the following
equivalence relation on $ \Om_\b$. We set
$\z \sim \tilde{\z} $ if for every $j\in \Z^d$,
\be
\label{G}
\int_{0}^{\b}\z_j (\t) d\t = \int_{0}^{\b}\tilde{\z}_j (\t) d\t .
\ee
For $y\in \R^{\Z^d }$, let $\Upsilon_\b (y) $ stand for the
equivalence class consisting of $\z$ such that
\be
\label{BC}
\b^{-1 } \int_{0}^{\b}\z_j (\t) d\t = y_j , \ \ \ j\in\Z^d .
\ee
We write $y \in \Upsilon_\b (y) $
 assuming that the former
$y$ stands for the constant loop $\om_j (\t ) = y_j $,
$j\in \Z^d $ and $\t \in [0, \b]$.
\begin{Pn}
\label{ACBpn}
For every $\nu \in {\cal G}^{{\rm qc}}(\b )$ and all
$B\in {\cal B}_\b $
\be
\label{NN}
\nu (B) = \nu ({\rm C}(B)) ,
\ee
i.e., every quasilocal Euclidean Gibbs state is supported on the
configurations consisting of constant loops.
\end{Pn}
Our first theorem establishes the relationship between
the families $\GCq$ and $\GCc$
\begin{Tm}
\label{MNtm}
For every $\nu \in \GCq$, there exists $\mu\in\GCc$,
such that
\be
\label{MN}
\mu(A) = \nu (B) = \nu ({\rm C}(B) ),
\ee
for all $A\in {\cal B}(\R^{\Z^d })$ and $B\in {\cal B}_\b $,
where ${\rm C}(B) \cong A$ in the sense of (\ref{A}). The mapping
$\nu \mapsto \mu $ (\ref{MN}) is a bijection.
\end{Tm}

In order to study the convergence of Euclidean Gibbs states
of the quantum model we shall use some notions
concerning the weak convergence of measures on metric
 spaces (see e.g. \cit{Pa}).
Consider a measure space $(X,{\cal B} (X))$, where $X$ is
a real separable metric space and $ {\cal B} (X)$ is
the Borel $\s$--algebra of its subsets.
Let ${\cal M}
(X)$ be the space
of all probability measures defined on ${\cal B} (X)$.
 Let $C_{{\rm b}} (X)$
stand for
the space of all bounded real valued continuous functions on $X$. The
topology on the space ${\cal M} (X)$ is defined by a system
of open
neighborhoods of a point $\mu \in {\cal M} (X)$, given as follows
\beq
\label{A8}
&  & V_\mu (f_1 , \dots , f_n ; \ve_1 , \dots , \ve_n )   =  \\
&  & \left\{ \nu \in {\cal M} (X) \vert \ \vert \int f_i d\nu -
\int f_i d\mu \vert < \ve_i ,\ i=1, \dots , n \right\} \nonumber
\eeq
with arbitrarily chosen $n \in \N$, $\ve_1 , \dots , \ve_n $ in $(0,
+\infty )$, and $f_1 , \dots , f_n $ in $C_{{\rm b}}(X)$. Such a
topology is said to be {\it the weak topology} on ${\cal M} (X)$.
If a net of measures $\{\mu_\a \}$ converges to a measure $\mu
\in {\cal M}(X)$ in this topology, we write $\mu_\a \Ra \mu$.
This convergence holds if and only if
$$
\int f d\mu_\a \ra \int f d\mu , \ \ \ \forall f\in C_{{\rm b}}(X).
$$

Now we may describe the weak convergence of the Euclidean
Gibbs measures when $m \ra +\infty $.

\begin{Tm}
\label{1Tm}
Let $\b>0 $, $\L \in \Lf$, and $y \in \R^{\Z^d} $ be chosen.
Then, for every $\z \in \Upsilon_{\b}( y ) $,
\be
\label{1602}
 \nu^{(m)}_{\b , \L} ( . \vert \z ) \Rightarrow
\nu_{\b , \L}^{{\rm qc}}(. \vert \z ) =
\nu_{\b , \L}^{{\rm qc}}(. \vert y ) , \ \ m \ra +\infty .
\ee
\end{Tm}
Unfortunately, this convergence does not imply the convergence
of the probability kernels defined by (\ref{12}), considered
as measures. In fact, for appropriate functions $f$, one has
from the above theorem
\beq
\label{220}
\int_{\Om_\b}f(\om )\pi_{\b, \L}^{(m)}(d\om \vert \z )
& = & \int_{\Om_{\b, \L}}
f(\om_\L \times \z_{\L^c} )\nu_{\b, \L}^{(m)}(d\om_\L \vert \z )
\\
& \ra & \int_{\Om_{\b, \L}}f(\om_\L \times \z_{\L^c} )
\nu_{\b, \L}^{\rm qc}(d\om_\L \vert y ), \ \ \z \in \Upsilon_\b (y),
\nonumber
\eeq
which shows that the dependence on $\z_{\L^c}$,
in contrast to (\ref{1602}), remains after
passing to the limit. Here we may prove only a somewhat weaker
result. Let $C_{{\rm b}} (\Om_\b )$ (resp. $C_{{\rm b}} (\R^{\Z^d})$)
stand for the set of all bounded continuous real valued functions on
$\Om_\b $ (resp. $ \R^{\Z^d}$) and
\be
\label{20}
\tilde{C}_{{\rm b}} (\Om_\b )
\ \sta \ \{ f\in \tilde{C}_{{\rm b}} (\Om_\b ) \ \vert
 \   f(\om ) = f(\tilde{\om}) \ \},
\ee
for every pair $\om \sim \tilde{\om}$ with the  equivalence
defined by (\ref{G}). One shows easily that for every $f\in
\tilde{C}_{{\rm b}} (\Om_\b )$ there exists $g\in C_{{\rm b}} (\R^{\Z^d})$
such that $f(\om ) = g(x)$ for $\om \in \Upsilon_\b (x)$.
The following theorem holds.
\begin{Tm}
\label{CCtm}
For every $f\in\tilde{C}_{{\rm b}} (\Om_\b )$ and any
$\b $, $\L\in \Lf$, and $\z \in \Upsilon_{\b}(y)$
$$
\int_{\Om_\b}f(\om )\pi_{\b, \L}^{(m)}(d\om \vert \z)
\ra
\int_{\R^{\Z^d}}g(x )\rho_{\b, \L}(dx \vert y)
$$
 when $m\ra + \infty $.
\end{Tm}

\begin{Rk}
\label{22rk}
Above we have restricted ourselves to the case of one--dimensional
oscillations of the particles. This was done only in order to avoid
further complications of notations and to make our considerations more
transparent. A generalization to the case of particles oscillating
in all directions ("vector case" where $x_k $ takes values in
some $\R^\nu $, $\nu >1$)
can be obtained with no additional troubles.
\end{Rk}

\section{Periodic Gibbs States and Order Parameters}

It is fairly well known
that, for $d\geq 2$, the models considered -- both quantum and
classical --
may undergo a phase transition when the inverse temperature $\b$
exceeds a certain value $\b_{*}$. The typical feature of this
phenomenon is the nonuniqueness of the Euclidean Gibbs
states. More precisely,
it may be proven (see e.g. \cit{AKRT}, \cit{AKKR}, \cit{PaY2}) that,
for the models considered here, the class of the called
tempered Gibbs measures (which are the Euclidean Gibbs states
with certain, physically motivated, restrictions on moments)
consists of exactly one element if the inverse
temperature $\b$ is small enough. In what follows,
the model
considered undergoes the phase transition if there exists $\b_{*}$
such
that for $\b > \b_{*}$, the class
of tempered Gibbs measures consists of more than one
element. This splitting of the class of tempered Gibbs measures,
which occurs when the inverse temperature $\b$ passes  $\b_{*}$,
is known as the phase transition in the model.
But in most nontrivial cases there are no possibilities to describe
the phase
transitions on this level. A much more realistic approach
is based on the use of {\it the order parameter}, which becomes
positive for $\b >\b_{*}$. The order parameter should describe the
symmetry breaking.
This means that the symmetry of the formal Hamiltonian, which is
inherited
by the unique, for $\b <\b_{*}$, tempered Gibbs measure, is no longer
proper, for $\b > \b_* $,
in the case where there exist more than one tempered
Gibbs measure.

Having the weak convergence of the Euclidean Gibbs measures of the
quantum model
to the Gibbs measure of the classical model we may study the possible
connections
between the order parameters in these models. For this purpose, the
most convenient
objects, of the type of those considered above, are the models
possessing
the translation invariance. Below we deal with translation
invariant models with the pair interaction,  described
by Hamiltonians of the type of (\ref{N1}) and (\ref{Cl1}). For
these models,
the translation invariance may be obtained if one assumes all $U_k $
being the same function $U$
and realizes  $J_{jk}$ as a
suitable function $J(.)$ of the
Euclidean distance $\vert j-k \vert $.
In order for local Gibbs measures to be invariant one may impose
periodic
boundary conditions instead of those established by means of
configurations
outside of $\L$. We shall now consider
this construction in more details.

Let a system of particles be  described by the following
formal translation invariant  Hamiltonians
\be
\label{t1}
H_{{\rm tran}} = \sum_{k }H_k^{(0)} + \sum_{k }
U(x_k ) + \half \sum_{j,k} J(\vert j - k \vert) x_j x_k ,
\ee
and
\be
\label{t2}
H_{{\rm tran}}^{{\rm cl }} = \sum_{k }\left( \half x_k^2 + U(x_k )
\right)
 + \half \sum_{j,k } J(\vert j - k \vert) x_j x_k ,
\ee
which correspond to the Hamiltonians (\ref{N1}) and (\ref{Cl1})
respectively. Here the Hamiltonian $H_k^{(0)}$ is
given by (\ref{2}) and all sums, as before, are taken over the whole
lattice $\Z^d$.
We also suppose that the function $J$ vanishes
when its argument exceeds some $r>0$,
and, in addition, $U$ and $J$
are assumed to obey the conditions (\ref{c}), (\ref{Qu2}). Consider
now
a box $\L\in \Lf $
\beq
\L = \{ k=(k_1 , \dots , k_d )\in \Z^d & \vert & k^{(0)}_l \leq
k_l \leq  k^{(1)}_{l}, \ \  l = 1,\dots , d \}, \nonumber
\eeq
$$
 k^{(0)}_l < k^{(1)}_{l}; \ \
k^{(0)}_l , k^{(1)}_l \in \Z^d \ .
$$
Given $\L$ and $ l= 1, \dots , d $, set
\be
\label{me1}
\vert j_l - k_l \vert_\L \ \sta \ {\rm min}\{\vert j_l - k_l \vert ;
\   k^{(1)}_l - k^{(0)}_l +1 -  \vert j_l - k_l \vert \ \},
\ee
and
\be
\label{me2}
\vert j - k \vert_\L^2 \ \sta \ \sum_{l=1}^{d}  \vert j_l - k_l
\vert_\L^2 ,
\ee
which defines the periodic metric in $\L$.
For this metric, one observes that
\be
\label{me3}
\vert j - k \vert_\L  \leq  \vert j - k \vert .
\ee
Chosen $\L$ and $\b$, we introduce the following
continuous real valued function on $\Om_{\b , \L}$
\be
\label{me4}
E^{{\rm per}}_{\b , \L} (\om_\L ) \ \sta \ \sum_{j\in \L}
\int_{0}^{ \b}
U(\om_j (\t )d\t + \half \sum_{j, k \in \L} \int_{0}^{\b}
J(\vert j - k \vert_\L ) \om_j (\t )\om_k (\t ) d\t ,
\ee
which will be used to construct the periodic local Gibbs
measures instead of the function (\ref{A1}).
Thereby we define
\be
\label{me9}
\nu_{\b , \L}^{{\rm per}}(d\om_{\L})
 \ \sta \  \frac{1}{Z_{\b,\L}^{{\rm per}}}\exp\{-
E_{\b,\L}^{{\rm per }}(\om_{\L}) \} \g_{\b , \L}^{(m)}(d\om_\L ),
\ee
Here
\be
\label{me11}
Z_{\b , \L}^{{\rm per}} \ \sta \ \int_{\Om_{\b , \L}}\exp\{ - E_{\b ,
\L}^{{\rm per}} (\om_{\L})\}\g_{\b ,\L }^{(m)}(d\om_\L ),
\ee
and the Gaussian measure  $\g_{\b ,\L }^{(m)}$
is the same as in (\ref{9}). Furthermore,
the quasiclassical periodic Gibbs measure is defined by (\ref{me9})
with $\g_{\b , \L}^{\rm qc}$ (\ref{Gcl})
instead of $\g_{\b , \L}^{(m)}$:
\be
\label{Dper}
\nu_{\b , \L}^{{\rm qcp}}(d\om_{\L})
 =  \frac{1}{Z_{\b,\L}^{{\rm qcp}}}\exp\{-
E_{\b,\L}^{{\rm per }}(\om_{\L}) \} \g_{\b , \L}^{\rm qc}(d\om_\L ),
\ee
\be
\label{me110}
Z_{\b , \L}^{{\rm qcp}} = \int_{\Om_{\b , \L}}\exp\{ - E_{\b ,
\L}^{{\rm per}} (\om_{\L})\}\g_{\b ,\L }^{qc}(d\om_\L ),
\ee

Then a version of Theorem
\ref{1Tm}, for the periodic Gibbs measures,
reads as follows.
\begin{Tm}
\label{6Tm}
Let $\b>0 $, $\L \in \Lf$  be chosen,
Then for the periodic Gibbs measures (\ref{me9}), (\ref{Dper}),
$$
\nu^{{\rm per}}_{\b , \L}  \Rightarrow
\nu_{\b , \L}^{{\rm qcp}} , \ \ \ m\ra +\infty .
$$

\end{Tm}

The classical analog
of (\ref{me4}) is
\be
\label{Ep}
I^{{\rm per}}_{ \L} (x_\L ) \ \sta \ \sum_{j\in \L}
U(x_j  + \half \sum_{j, k \in \L}
J(\vert j - k \vert_\L ) x_j x_k  ,
\ee
which we use to construct the classical periodic Gibbs
measure
\be
\label{meA5}
\mu_{\b , \L}^{{\rm per}} (dx_\L ) =
\frac{1}{Y_{\b , \L}^{{\rm per}}}
\exp \{ - \b I_{ \L}^{{\rm per}} (x_\L )\}
\chi_{\b ,\L} (dx_\L ),
\ee
where
\be
\label{meY}
Y_{\b , \L}^{{\rm per}} = \int_{\R^{\L}}
\exp \{ - \b I_{ \L}^{{\rm per}} (x_\L  ) \}
\chi_{\b , \L} (dx_\L ),
\ee
and the Gaussian measure  $\chi_{\b ,\L }$
is defined by (\ref{A6}), (\ref{A7}).
 Now we introduce the order parameters
which become positive for $\b > \b_{*}$, manifesting
the appearance of the long range order. From now on,
in addition to the previous assumptions,
we assume that the anharmonic potential $U$ is an even function,
which means that the symmetry being broken is $Z_2 $.
Then the corresponding order parameters are defined by means
of the following moments of the periodic Gibbs
measures
\beq
\label{OP1}
 P_\L (m) & \sta & \int_{\Omega_{\b , \L}}
\left(\frac{1}{\vert \L \vert} \int_{0}^{ \b }\sum_{j\in \L}
\om_j (\tau)d\tau \right)^2 \nu^{{\rm per}}_{\b , \L}(d\om ), \\
\label{OP2}
 Q_\L & \sta & \int_{\R^{ \L}}
\left(\frac{1}{\vert \L \vert} \sum_{j\in \L}
x_j  \right)^2 \mu_{\b , \L}^{{\rm per}}(dx_\L ).
\eeq
Here $\vert \L \vert$ stands for the cardinality
of $\L$. In this case Theorem \ref{6Tm} implies.
\begin{Co}
\label{OPco}
Let $\b>0 $, $\L \in \Lf$  be chosen, then
\be
\label{OP5}
\lim_{m \ra +\infty }P_\L (m) = Q_\L .
\ee
\end{Co}
Another relation between
the moments $P_\L (m)$ and $Q_\L$ may be established for
a special choice of the function $U$ and under
additional conditions imposed on the function $J$.
\begin{Pn}
\label{P1Pn}
For the considered quantum model, let  $J$ be
a nonnegative monotone decreasing function
and $U$ have the following form
\be
\label{Phi4}
U (x_j ) = ax_j^2 + \sum_{l=2}^{p}b_l x_j^{2l} , \ \ a\in \R ,
\ b_p >0, \ b_l \geq 0, \ l = 2, \dots , p, \  p\geq 2.
\ee
Then for every $\L \in \Lf $, the moment $ P_\L (m) $ is a monotone
increasing function of $m \in (0, +\infty)$, i.e.,
for arbitrary $ m' > m$,
$$
P_\L (m') \geq P_\L (m).
$$
\end{Pn}
The proof of this assertion is based
on the properties of $J$, $U$,
it will be done in a separate work \cit{AKK2}.

Now we define  (see e.g. \cit{Ko})
\beq
\label{OP3}
P(m) & \sta & \lim_{\L\nearrow \Z^d } P_\L (m) \\
\label{OP4}
Q & \sta & \lim_{\L\nearrow \Z^d } Q_\L,
\eeq
which are the order parameters for the translation invariant
quantum and classical models respectively.
 In fact, in order to prove the appearance of the long
range order one does not need to find
these limits explicitly. It is enough to show that the sequences
$\{P_\L (m)\}$, $\{Q_\L \}$ are uniformly,
with respect to $\L$, below bounded.
Combining these relations one concludes that in this case,
the uniform boundedness, when $\L \nearrow \Z^d $, of the sequence
$$
 P_\L (m) \geq p(m) >0,
$$
implies
the appearance of long range order
for all $m' > m $, as well as for $m = +\infty$,
that means in the classical model.

\section {The Proofs}

The proof of all our theorems is based on the following lemma,
which is proven in the final part of this section.
\begin{Lm}
\label{1Lm}
For every $\L \in \Lf$, $\b > 0$,
$ \g_{\b , \L }^{(m)}  \Rightarrow \g_{\b , \L }^{\rm cl}$.
\end{Lm}
{\bf Proof of Proposition \ref{ACBpn}.}
We set
\be
\label{Cyl}
{\cal C} \ \sta  \ \{B = B_\L \times \Om_{\b , \L^c} \
\vert \ B_\L \in {\cal B}_{\b , \L}, \ \L \in \Lf \}.
\ee
\be
\label{Cycl}
{\cal C}_{{\rm cl}} \ \sta  \ \{A = A_\L \times \R^{ \L^c} \
\vert \ A_\L \in {\cal B}(\R^{\L}), \ \L \in \Lf \}.
\ee
By the definition of the probability kernels (\ref{12}),
which is also valid for the quasiclassical ones,
\be
\label{Cy2}
\pi_{\b,\L}^{{\rm qc}}(B_\L \times \Om_{\b , \L^c}\vert \om)
=\nu_{\b,\L}^{{\rm qc}}(B_\L \vert \om_{ \L^c}).
\ee
Since $\nu $ is in ${\cal G}^{{\rm qc}}(\b)$ it obeys the
equilibrium equation (\ref{14}) with the quasiclassical kernels
$\pi_{\b,\L}^{{\rm qc}}$. Let some $B\in {\cal C}$ be chosen.
Then it is a cylinder $B_\L \times \Om_{\b , \L^c}$ with certain
$\L\in \Lf$, thus one can choose in (\ref{14}) this $\L$.
This and (\ref{Cy2}) yield
\beq
\label{Cy3}
& & \nu (B_\L \times \Om_{\b , \L^c})  =  \int_{\Om_\b}
\pi_{\b,\L}^{{\rm qc}}(B_\L \times \Om_{\b , \L^c}\vert \om)
\nu (d\om ) \\
& = & \int_{\Om_\b}\nu_{\b,\L}^{{\rm qc}}
(B_\L \vert \om_{ \L^c})\nu(d\om)
 =  \int_{\Om_\b}\nu_{\b,\L}^{{\rm qc}}
({\rm C}(B_\L ) \vert \om_{ \L^c})\nu(d\om),
\nonumber
\eeq
which follows from (\ref{Ng}). Thus
$$
 \nu (B_\L \times \Om_{\b , \L^c}) =
 \nu ({\rm C}(B_\L ) \times \Om_{\b , \L^c}) , \ \ \L\in \Lf .
$$
This implies (\ref{NN})
\kasten

{\bf Proof of Theorem \ref{MNtm}.}
Comparing (\ref{A1}) and (\ref{E}) one concludes that for
every $\z \in \Upsilon_\b (y) $ and
$\om_\L \in \Om_{\b, \L}^{{\rm qc}}$, such that $\om_k (\t )=
x_k $, $k\in \L$
\be
\label{Cy4}
E_{\b, \L } (\om_\L \vert \z ) = \b I_{\L} (x_\L \vert y).
\ee
This and (\ref{Ng}) imply for such $\z$
\be
\label{221}
\nu_{\b,\L}^{{\rm qc}}(B_\L \vert \z) =
\nu_{\b,\L}^{{\rm qc}}({\rm C}(B_\L ) \vert \z) =
 \mu_{\b,\L}(A_\L  \vert y),
\ee
where ${\rm C}(B_\L ) \cong A$.
Now let us define on ${\cal C}_{{\rm cl}}$ the following cylinder
measure
\be
\label{Cy44}
\mu(A) = \mu (A_\L \times \R^{\L^c }) \ \sta \
 \nu ({\rm C}(B_\L ) \times \Om_{\b , \L^c}), \ \ \
{\rm C}(B_\L ) \cong A .
\ee
Then
\beq
\label{Cy45}
\mu(A) & = & \mu (A_\L \times \R^{\L^c }) =
\int_{\Om_\b}\nu_{\b,\L}^{{\rm qc}}
({\rm C}(B_\L ) \vert \om_{ \L^c})\nu(d\om) \\
&=&
\int_{\Om_\b^{\rm qc}}\nu_{\b,\L}^{{\rm qc}}
({\rm C}(B_\L ) \vert \om_{ \L^c})\nu(d\om)
=
\int_{\R^{\Z^d}}\mu_{\b,\L}
(A_\L \vert x_{ \L^c})\nu(dx). \nonumber
\eeq
Here we have taken into account that the measure $\nu $ has
$\Om_\b^{{\rm qc}}$ as support.
Since $\mu$ is defined by a measure, it can be continued
as a measure on the whole $\s$-algebra ${\cal B} (\R^{\Z^d})$.
Directly from (\ref{Cy45}) one sees
 that this measure obeys
the equilibrium equation (\ref{cl14}) thus it belongs to
${\cal G}^{{\rm cl}}$. Now for every $\mu \in {\cal G}^{{\rm cl}}$,
one can define a cylinder measure on ${\cal C}$
as given  by (\ref{Cyl}) by
a relation of the type of (\ref{Cy44}) and repeat the above
steps obtaining an element of ${\cal G}^{{\rm qc}}$.
\kasten


{\bf Proof of Theorem \ref{1Tm}.}
We remind that in the case considered the function
$E_{\b , \L}(\om_\L \vert \z )$ is given by (\ref{A1}).
Then the density
$$
F_{\b ,\L}(\om_\L \vert \z ) \ \sta  \
\frac{\nu^{(m)}_{\b , \L}(d\om_\L \vert
\z )}{\g_{\b,
\L}^{(m)}(d\om_\L )}
$$

 may be written as
$$
F_{\b ,\L}(\om_\L \vert \z ) = \frac{1}{Z_{\b , \L }(\z)}
\exp\left\{ -
\sum_{j \in \L, k \in \L^{c}}
J_{jk}\int_{0}^{\b }\om_j (\t) \z_k (\t)d\t \right\}\Psi_{\b,
\L}(\om_\L ),
$$
with a certain $\Psi_{\b , \L} \in C_{{\rm b}} (\Om_{\b , \L})$.
Therefore, for an arbitrary function $G\in C_{{\rm b}}(\Om_{\b , \L})$,
one has
\beq
\label{v}
& & \int_{\Om_{\b , \L}}G(\om_\L ) \nu^{(m)}_{\b , \L}(d\om_\L \vert
\z )
= \frac{1}{Z_{\b , \L }(\z)}
\int_{\Om_{\b , \L}} G(\om_{\L})\Psi_{\b , \L }(\om_\L ) \nonumber \\
&  & \exp\left\{-\sum_{j \in \L, k \in \L^{c}}
J_{jk}\int_{0}^{\b }\om_j (\t) \z_k (\t)d\t \right\}\g_{\b,
\L}^{(m)}(d\om_\L ) \nonumber \\
& \lra & \frac{1}{Z_{\b , \L }(\z)}
\int_{\Om_{\b , \L}} G(\om_{\L})\Psi_{\b , \L }(\om_\L ) \nonumber \\
&  & \exp\left\{
- \sum_{j \in \L, k \in \L^{c}}
J_{jk}\int_{0}^{ \b }\om_j (\t) \z_k (\t)d\t \right\}
\g^{{\rm qc}}_{\b, \L}(d\om_\L ) \nonumber \\
 & = & \frac{1}{Z_{\b , \L }(\z)}
\int_{\R^\L} G(x_{\L})\Psi_{\b , \L }(x_\L )
 \exp\left\{ -\sum_{j \in \L, k \in \L^{c}}
J_{jk}x_j  \int_{0}^{ \b }\z_k (\t)d\t \right\}
\chi_{\b, \L}(dx_\L )
\nonumber \\
 & = & \frac{1}{Z_{\b , \L }(\z)}
\int_{\Om_{\b , \L}} G(\om_{\L})\Psi_{\b , \L }(\om_\L )
 \exp\left\{ -\sum_{j \in \L, k \in \L^{c}}
\b J_{jk}\om_j (0) y_k \right\}
\g^{{\rm qc}}_{\b, \L}(d\om_\L )
\nonumber \\
& = & \int_{\R^\L}G^{{\rm cl}}(x_\L)\mu_{\b ,\L}
(d\om_\L \vert \xi ) , \nonumber
\eeq
where $G^{{\rm qc}}$ is a restriction of $G$ on
$\R^\L \cong \Om_{\b , \L}^{{\rm qc}} $ and
we have taken into account that $\z\in \Upsilon_\b (y)$.
\kasten

The proof of Theorem \ref{6Tm} may be performed
by a repetition of the arguments just used.

{\bf Proof of Theorem \ref {CCtm}. }
It follows directly from (\ref{220}) and (\ref{221}).
\kasten

Now it remains to prove
Lemma \ref{1Lm}. To this end we use the following known property
of Gaussian
measures on a Hilbert space ${\cal H}$ (see e.g. pp. 153--155 of  book
\cit{Pa}).
\begin{Pn}
\label{1Pr}
Let a net of Gaussian measures $\{ \g_\a \}$ on a separable real
Hilbert space ${\cal H}$ be given. Let also each measure $\g_\a $
have zero mean and a trace
class operator on ${\cal H}$,
 $S_\a $, as a covariance operator. Suppose that the net $\{ S_\a \}$
converges in the trace norm to an operator $S$. Then there exists
a Gaussian measure $\g$ on the space ${\cal H}$ such that
the operator $S$ is its covariance operator, and $\g_a \Ra \g$.
\end{Pn}
{\bf Proof of Lemma \ref{1Lm}.} First of all we construct explicitly
the covariance operators of the Gaussian measures $\g_{\b , \L}$
and $\g_{\b , \L}^{{\rm qc}}$ defined by (\ref{0c}) and
(\ref{Gcl}) respectively.
The former one implies
\be
\label{0d}
S_{\b , \L}(m) = \sum_{j\in \L}  S_{\b , j} (m)
P_{\b , j},
\ee
where $P_{\b , j}$ is the projector from
${\cal H}_{\b , \L}$ onto the space
${\cal H}_{\b , j}$. Here, as before, we  omit the subscript
$j$ when this does not cause any ambiguities.

In the sequel we will need a base of the Hilbert space $\Hb =
{\cal H}_{\b , j}$, which we choose as the following
orthonormal set the eigenfunctions of $\D_\b $
\beq
\label{HH2}
e_{q}(\t) & = & \sqrt{\frac{2}{\b}}\cos q\t , \ {\rm for} \ q>0;
 \ \ \ \ q\in {\cal Q}
\ \sta \ \{\frac{2\pi}{\b}n \ \vert \ n \in \Z \},  \nonumber \\
e_{q}(\t) & = & \sqrt{\frac{2}{\b}}\sin q\t , \  {\rm for} \ q<0;
\ \ \ \ \ e_{0}(\t) =
\sqrt{\frac{1}{\b}}.
\eeq
The operator $S_{\b , \L}(m)$ acts on $ {\cal H}_{\b , \L}$
as a positive compact operator,
 hence it has the
canonical representation:
\be
\label{0e}
S_{\b , \L}(m) = \sum_{j\in \L}
\left(\sum_{q\in {\cal Q}}(mq^2 +1)^{-1}W_q \right)
P_{\b , j}.
\ee
where $W_q $ is the projector from ${\cal H}_{\b}$ onto the direction
$e_q $ (see (\ref{HH2})).
Having this representation we prove Lemma \ref{1Lm} by showing
that the net of covariance operators
$\{ S_{\b ,\L}(m) \vert m \in (0, +\infty) \}$ converges in the
trace norm to the covariance operator of the measure
$\g^{\rm qc}_{\b,  \L}$ defined by (\ref{Gcl}), (\ref{A6}),
(\ref{A7}).
Let us construct the covariance operator
of the latter measure. To this end
we write its Fourier transformation, which should have
the form (\ref{5}), valid for all
Gaussian measures:
\beq
\label{F1}
\Gamma^{{\rm qc}}_{\b,  \L} (\vp_\L) & \sta & \int_{{\cal H}_{\b ,
\L}}
\exp\{i(\vp_\L , \om_\L )_{\b , \L}\}
\g^{{\rm qc}}_{\b,  \L} (d\om_\L ) \\
& = & \exp\{ - \frac{1}{2} (\vp_\L , S_{\b ,\L}^{{\rm qc}}
\vp_\L )_{\b , \L} \},
\eeq
where the scalar product  $(. , . )_{\b , \L}$ is defined by
(\ref{3ab}). On the other hand, (\ref{Gcl}) implies that
the measure  $\g^{\rm qc}_{\b,  \L}$
 is supported on the subset $\Om^{{\rm qc}}_{\b , \L} \subset
\Om_{\b , \L} \subset {\cal H}_{\b , \L} $, where it coincides
with the measure $\chi_{\b , \L}$
given by  (\ref{A6}), (\ref{A7}). This
yields in (\ref{F1})
\beq
\label{F2}
  \Gamma^{{\rm qc}}_{\b,  \L} (\vp_\L )
& = &\left( \frac{\b}{2\pi}\right)^{ \frac{ \vert \L \vert }{2}}
  \int_{{\R}^{\L }}\exp \left\{ i\sum_{j\in \L }
 x_j \int_{0}^{\b}\vp_j (\t)d\t \right\} \nonumber \\
&  &  \exp\left\{ -\frac{\b}{2}
 \sum_{j\in \L } x_{j}^2 \right\}\prod_{j\in \L }
 dx_{j}  \nonumber  \\
& &   = \exp\left\{ - \frac{1}{2\b}\sum_{j\in \L }
\left( \int_{0}^{\b} \vp_j (\t)d\t \right)^2 \right\} \nonumber \\
&  &  = \exp\left\{ -\frac{1}{2}\sum_{j\in \L }
 (e_0 ,\vp_j )_{\b}^2 \right\}.
\eeq
Here we have used the eigenfunction $e_0$ given by (\ref{HH2}).
Comparing the latter form of $\Gamma_{\b , \L}^{{\rm qc}}$
with the definition (\ref{F1}), one concludes that
\be
\label{F3}
S_{\b , \L}^{{\rm qc}} = \sum_{ j\in\L} W_0
P_{\b , j},
\ee
where $W_0$ is a projector in ${\cal H}_{\b }$ on the
direction $e_0$.
Now we use the canonical representation (\ref{0e}) and obtain
$$
S_{\b , \L}(m) - S_{\b , \L}^{{\rm qc}} =
\sum_{j \in \L} \left( \sum_{q\in {\cal Q}\setminus \{0\}}
\frac{1}{mq^2 +1}W_q \right)P_{\b ,j} ,
$$
which yields
\beq
\label{F4}
{\rm trace}(
S_{\b , \L}(m) - S_{\b , \L}^{{\rm qc}}) & = &  \vert \L \vert
\sum_{q\in {\cal Q}\setminus \{0\}}\frac{1}{mq^2 +1}
 \leq   \vert \L \vert
\sum_{q\in {\cal Q}\setminus \{0\}}\frac{1}{mq^2} \nonumber \\
& = &
\frac{1}{m} \left( \frac{ \vert \L \vert \b^2}{2\pi^2} \sum_{n \in \N}
\frac{1}{n^2}\right) \lra 0 , \ \ m \ra  + \infty .    \nonumber
\eeq
\kasten

{\bf Acknowledgment} Yuri Kondratiev is grateful for the support
through the INTAS project 94-0378. Yuri Kozitsky is grateful
for the kind hospitality in Bochum, where
 this work was brought into the final form, and for the
financial support of his stay from
SFB--237 (Essen-Bochum-D{\"u}sseldorf).

\end{document}